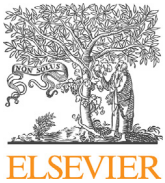
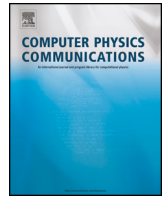

# Distributed simulation and visualization of the ALICE detector magnetic field ☆

Piotr Nowakowski [a],*, Przemysław Rokita [a], Łukasz Graczykowski [b]

[a] *Faculty of Electronics and Information Technology, Warsaw University of Technology, Nowowiejska 15/19, 00-665 Warsaw, Poland*
[b] *Faculty of Physics, Warsaw University of Technology, Koszykowa 75, 00-662 Warsaw, Poland*

### A B S T R A C T

The ALICE detector at CERN uses properties of the magnetic field acting on charged particles as part of the particle tracking and identification system – via measuring the strength of bending of charged particles in a magnetic field supplied by electromagnets. During the calibration of the detector the magnetic field was measured by the scientific team. The measurement points were then fitted using Chebyshev polynomials to create a field model. That model is used extensively in particle trajectory reconstruction, but for maximum compatibility it is handled by CPU-only software. We propose multiple approaches to the implementation of the model in a shader language, which allows it to run in a graphical GPU environment. This makes the model run more efficiently, as well as enables it to be used for field visualisation, which was not previously done for ALICE.

## 1. Introduction

**ALICE** (**A** **L**arge **I**on **C**ollider **E**xperiment) [1] is one of the detector experiments located at the Large Hadron Collider (LHC) [9], primarily dedicated to the study of ultrarelativistic heavy-ion collisions. The main objective of ALICE is the study of the properties of the produced deconfined state of matter, the quark–gluon plasma (QGP) which, according to theory, existed few μs after the creation of our Universe [2,14,13,18]. A number of observables are studied to characterise the properties of the hot and dense matter produced in the laboratory in energetic collisions of lead nuclei (Pb–Pb) and systematically compare them with proton-proton (pp) collisions [10,11].

ALICE is composed of several detection systems, which are embedded in the magnetic field generated by a solenoid and a dipole magnet. The magnetic field is used to bend the trajectory of charged particles in order to measure their momenta.

Characteristics of the fields generated by both magnets were measured in 2005 by the ALICE Collaboration. The collected data was then used to create a model of the magnetic field for *AliROOT*, the experiment's software framework used in physics calculations. The public GitHub repository[1] contains the code of the framework as well as the magnetic field model data. The modelling method (Chebyshev polynomials) and the implementation itself was chosen by authors to run efficiently on the CPU, as the framework is designed to run on any machine (and as such it was not possible to assume access to optional features such as SIMD processor instruction set extensions or external accelerators such as a GPU). The model is most commonly used in track reconstruction.

3D visualisation tools of ALICE data (i.e. Event Display [16] and MasterClass software suite [12]) are designed to display reconstructed tracks; however, they do not use this accurate field model, instead implementing the simplest possible — uniform — approximation of only the solenoid field (the dipole field is not taken into account at all).

The main objective of this paper is to create a novel version of the model implementation capable of running on the GPU in a graphical (OpenGL) scenario to create a possibility of using the magnetic field model in real-time, 3D visualisations of particle collisions, and for displaying the field itself to showcase how different parts of the detector machinery work.

We have examined two main approaches of using the field model in a GPU context: (a) *Look Up Table* based, where the magnetic field samples (obtained by querying the original model) are, at least partially, stored in the GPU's Video RAM and are used to evaluate the field; and (b) *Equation* based, in which the original algorithm, adapted to the GLSL shading language and GPU programming environment, is used. All variations of the algorithm were tested for performance and memory requirements, as well as for accuracy with the results obtained from the original *AliROOT*

---

implementation and with the uniform field approach used in the mentioned visualisation software. The development version of all implemented variations of the algorithm and of the test program can be obtained from the GitHub repository.[2]

Performance improvements to the original *AliROOT* algorithm were attempted in [22], although the GPU approach was not explored in this work (scope limited to CPU, like in the original framework). In preparation for the upcoming period of data-taking, which will operate on upgraded hardware components of many detectors (dubbed "LHC Run 3"), a new software framework was developed for ALICE — $O^2$ [19,4]. In this framework a new model was created that can run on both the CPU and GPU, although it supports only the Solenoid magnet volume in general, as well as allows the user to work only in a single fragment of the whole volume at once, which has to be selected before computations can be made. The implementation is also targeted for the compute (CUDA / OpenCL / HIP) environment, so it can not be used as-is for generating graphics with OpenGL.

The rest of this paper is structured as follows. In Section 2 a brief introduction to the ALICE experiment is presented along with the coordinate systems used in the magnetic field model. In Section 3 the original field algorithm from *AliROOT* is explained in detail. Section 4 presents all implemented approaches of using the field on the GPU with their up- and downsides listed. In Section 5 we explain how the implemented methods were tested against each other and against the original *AliROOT* algorithm. Section 6 presents the experimental results. Finally, Section 7 contains the conclusion drawn from the results, as well as ideas for further research.

## 2. ALICE detector

The ALICE experiment is composed of two main parts (see Fig. 1). The first one is the so-called "central barrel", centred at the interaction point and placed inside a large solenoid magnet (in red). It contains detectors designed to record particle trajectories and identification over a broad momentum range via a number of techniques. The main tracking devices are the Inner Tracking System (ITS) [7], composed of six cylindrical layers of silicon detectors and the Time Projection Chamber (TPC) [8], which is a large (volume of 88 m$^3$) cylindrical detector filled with mixture of a noble gas and $CO_2$. It is divided into two halves split by the central electrode. At the end of each half there is a readout plane composed of 18 sectors, covering full azimuthal angle. Each of those sectors contains 159 pad rows which are arranged in radial orientation. The curvature of the particle trajectory is obtained mainly from the signal registered by the TPC. The second part, so-called "muon arm", is located at the forward region of the central barrel. It contains the dimuon spectrometer [3], embedded inside a dipole magnet, with the aim to record trajectories of muon pairs.

### 2.1. Coordinate systems

The magnetic field model (see Section 3) uses two coordinate systems: a Cartesian system and a cylindrical system. The Cartesian coordinate system has its origin in the centre of the barrel, where the accelerated particle beams from the LHC cross each other to allow particle collisions to occur. This point is called the Interaction Point (IP). The $z$ is along the LHC beam axis and points in a direction from the muon spectrometer to the central barrel. The $y$ axis points up. The $x$ axis is pre-defined based on the other two to complete a right-handed system.

---

[2] Developed software repository (Accessed: 2021-08-15) — https://github.com/pnwkw/distributed_field.

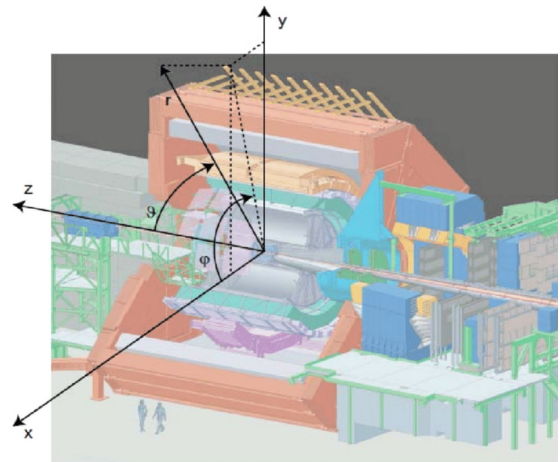

**Fig. 1.** The ALICE detector (central barrel and the muon spectrometer) and its coordinate systems [6]. (For interpretation of the colours in the figure(s), the reader is referred to the web version of this article.)

The cylindrical coordinate system uses the same origin point and the $z$ axis as the Cartesian system. The $r$ coordinate is the Euclidean distance of a point from the origin. The angle $\phi$ is the angle between a point projected on the $x\ y$ plane and the $x$ axis (see Fig. 1).

## 3. Magnetic field model

The algorithm in *AliROOT* is designed to offer a value of the field vector $\vec{B}$ (with precision depending on the location, but in range of $2 \times 10^{-4}$ to $5 \times 10^{-3}$ kGauss) in the ALICE detector volume (12 m × 11 m × 36 m). The desired position for evaluation should be expressed in the ALICE Cartesian coordinate system. The input position is expressed in centimetres (the maximal spatial precision for the model).

The magnetic field evaluation is constrained in [-1760 cm, 850 cm] on the $z$ axis — outside this range the returned field vector is zero. Influence of both magnets is modelled separately and the correct model is chosen according to the position on the same axis. This creates two cases:

- $z \leq -550$ cm, where a model for the Dipole magnet is used,
- $z > -550$ cm, where a model for the Solenoid magnet is used.

### 3.1. Dipole section

The magnetic field model of the Dipole magnet consists of many parametrisation segments. For each segment a unique Chebyshev polynomial (of varying order) has been assigned which is used to calculate the final field value using Cartesian ($x$, $y$, $z$) coordinates (see Fig. 1) of the requested point. Segments for this model have cuboid shapes. A visualisation of the segment structure can be seen in Fig. 2. To fetch the necessary segment for evaluation, the algorithm:

1. checks if the previously cached segment can be used for the current point,
2. if it matches, the corresponding polynomial is evaluated and the result returned,
3. if not, the correct row of segments on the $z$ axis is found using binary search,
4. then the correct row of segments on the $y$ axis is found using linear search,
5. finally, the correct segment is found on the $x$ axis using the linear search again.

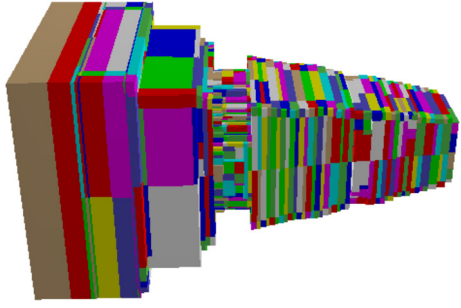

**Fig. 2.** Visualisation of parametrization segments for the dipole magnet (1482 segments in total) [22].

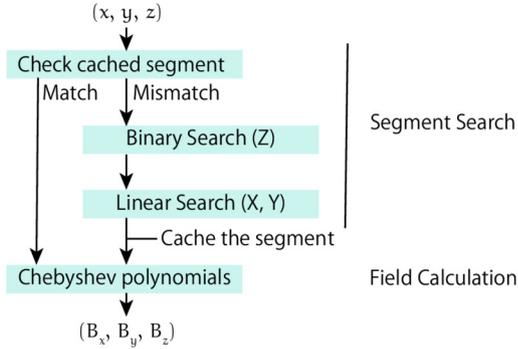

**Fig. 3.** Schematic of the field evaluation for the Dipole magnet model [22].

6. The newly found segment replaces the previous one in cache, and the corresponding polynomial is evaluated and the result returned.

A schematic for the field evaluation algorithm can be seen in Fig. 3. The cache improves performance when consecutive evaluation requests lie inside the same segment, which is often the case in the main use case of the algorithm — track reconstruction.

### 3.2. Solenoid section

The magnetic field model of the Solenoid magnet is also split into parametrisation segments. To take advantage of the circular symmetry of the magnet, those segments are arranged in a cylindrical coordinate system $(r,\phi,z)$ (see Fig. 1) — they have a ring fragment shape instead of a cuboid. The fetch algorithm is identical except for the fact, that the coordinates of the requested point have to be converted to the cylindrical form first, the search is done on $z$, $\phi$, $r$ (in this order) instead, and the evaluated field value (as it is expressed in the cylindrical system as well) has to be converted back to the Cartesian before it is returned to the user.

### 3.3. Field evaluation

In the *AliROOT* algorithm, after the correct model segment is identified, a mapping from the requested point $\vec{P}$ (as-is in Cartesian coordinates in Dipole case, converted to Cylindrical coordinates in the Solenoid case) to "internal" representation $\vec{P_{int}}$ is performed. The mapping is done in such a way that every component of the $\vec{P_{int}}$ is in range $[-1, 1]$, suitable for use as the parameter for a Chebyshev polynomial. $\vec{P_{int}}$ is obtained using the formula $\vec{P_{int}} = (\vec{P} - \vec{offsets}) \cdot \vec{scales}$ ($\cdot$ is element wise multiplication), where *offsets* and *scales* are three dimensional vectors unique to each segment.

In the Dipole case, each component of the resulting magnetic field vector $\vec{B}$ is calculated using a different Chebyshev polynomial (to avoid label confusion referenced here as $B^0$, $B^1$, $B^2$), using the $x$ component of the position vector as the parameter. Coefficients of these polynomials (referenced here as $c_i$) are not stored in the segment data directly. Instead, they are calculated on the fly using a different set of Chebyshev polynomials, this time evaluated using the $y$ component of the position vector. Coefficients of these polynomials (referenced here as $c_{ij}$) are not stored in the segment data either — they are calculated on the fly by a final set of Chebyshev polynomials, evaluated using the $z$ component of the position vector. Coefficients of these polynomials are stored in the per segment data (referenced here as $c_{ijk}$).

For example, to calculate the first component $\vec{B}[0]$ of the resulting field vector $\vec{B}$, the following operations have to be performed:

$$c_{ij}^0 = \sum_{k=0}^{K_j^0} c_{ijk}^0 T_k(z),$$

$$c_i^0 = \sum_{j=0}^{J_i^0} c_{ij}^0 T_j(y),$$

$$B^0 = \sum_{i=0}^{I^0} c_i^0 T_i(x),$$

$$\vec{B}[0] = B^0.$$

$T_n$ function is a single Chebyshev polynomial of $n$-th order. $K$, $J$, and $I$ are the maximum order of the polynomial in parametrization, which can be different for every single calculated coefficient.

The same set of operations (although with a different number and values for the coefficients) has to be performed to obtain $B^1$ and $B^2$, completing the calculation of the $\vec{B}$ vector for the requested point in space.

In the Solenoid case the steps required to calculate the $\vec{B}$ are identical, except the input vector is in Cylindrical coordinates, so $(r, \phi, z)$ are used instead of $(x, y, z)$. As mentioned before, this results in the calculated $\vec{B}$ vector to be in the Cylindrical coordinate system as well, so the final step requires a conversion of this result back to the Cartesian system.

## 4. GPU implementation

In total, six algorithms of magnetic field data access were developed:

- Shader Storage Buffer,
- 3D Texture,
- Sparse 3D Texture,
- GLSL (reimplementation of the original algorithm in a shader) with segment cache,
- GLSL without segment cache,
- Constant.

The following subsections will explain in greater detail each of the technique used, as well as advantages and disadvantages of each of them.

### 4.1. Shader storage buffer

The spatial resolution of the model is not unlimited (see Section 3), therefore the most straightforward way of using it in a GPU is to keep a pre-computed array of samples (3D vectors) in its memory. Unfortunately storing the whole detector volume is not feasible due to large memory requirements — around 2700 million `float` vectors (12 B each) — which accounts of 32 GiB of data in total.

P. Nowakowski, P. Rokita and Ł. Graczykowski  Computer Physics Communications 271 (2022) 108206

The space problem can be somewhat offset by decimating the input data, although this will cause discrepancies with the original model (see Section 6 for accuracy comparisons with the *AliROOT* algorithm). As an example, keeping every fourth sample in each dimension will reduce the amount of space needed by a factor of 64, to around 500 MiB.

In OpenGL 4.6, there are two ways of providing shaders with user-defined data:

- Uniform Buffer,
- Shader Storage Buffer.

Unfortunately the (potentially faster) Uniform cannot be used due to very small amount of data it can hold (16 KiB). The size guaranteed by the OpenGL Specification for the Shader Storage Buffer is 128 MiB, although graphics cards usually allow allocating almost all of the memory available for this kind of buffer,[3] making it our only choice.

There are also restrictions on the address boundaries of variables stored in such a buffer, which affect their possible layout in memory. The std430 offers the tightest possible memory arrangement, but vectors stored in a buffer must at least be aligned to 16 B boundaries. Our case of 3D vectors requires either a manual management of individual vector components (which would slow down the program) or rounding up to 4D, which further increases the memory usage — from around 500 MiB to around 656 MiB in the mentioned example.

The field's data, uploaded to the GPU, is arranged as a linearised 3D array. With the width, height and depth of the array known at compile time, an arbitrary 3D point inside the detector volume can be transformed into an index into the array with a simple calculation, done by the world_to_index function (see Fig. 4). The ALICE coordinate system is centred on the Interaction Point, which is inefficient for the field storage, as the detector is not symmetrical. We use a slightly different system (where the origin is in the geometrical centre of the detector), which requires a simple translation (realised by the offsets global variable). Then, scales global variable is used to obtain the requested point's position after the sample decimation. Finally, to calculate a linear array index from a coordinate, the size of the detector in each dimension is required — this information is stored in the detector_dimensions_scaled global variable.

### 4.2. 3D texture

If 3D textures of sufficient size (equal to the size of the detector volume after sample decimation) are supported by the graphics card, it is possible to store the field samples in a texture image. OpenGL offers GL_RGB_32F format, which can store the field vector samples tightly, with no padding — for the example decimation level, such texture takes around 500 MiB of storage space.

Using a 3D texture has some advantages over a Shader Storage Buffer, such as faster data access if the texture is handled by a dedicated hardware on the GPU (a texture unit) and a built-in linear interpolation between samples. Because the field is continuous inside the detector, applying this interpolation to the data should improve accuracy, if compared with the previous method. A sample slice of the 3D texture, with the field vectors interpreted as RGB data, can be seen in Fig. 5.

An arbitrary 3D point inside the detector volume can be very easily transformed into a Cartesian texture ($u$, $v$, $w$) coordinate,

---

[3] Shader Storage Buffer Object extension specification (Accessed: 2021-08-15) — https://www.khronos.org/registry/OpenGL/extensions/ARB/ARB_shader_storage_buffer_object.txt.

```
int world_to_index(vec3 position) {
    ivec3 r = ivec3((position.zxy + offsets) / scales);

    return (r.p * detector_dimensions_scaled.s *
    detector_dimensions_scaled.t) +
    r.t * detector_dimensions_scaled.s + r.s;
}

vec3 Field(vec3 pos) {
    int index = world_to_index(pos);
    return data[index].xyz;
}
```

**Fig. 4.** Field access shader code for the Shader Storage Buffer implementation.

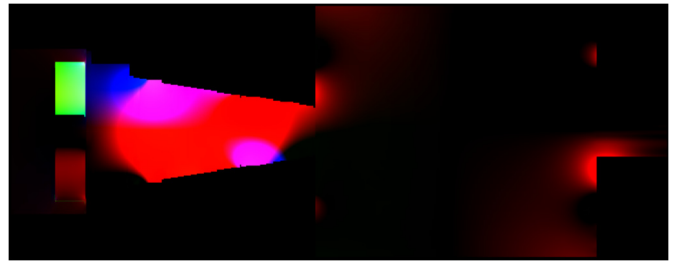

**Fig. 5.** A sample slice of the 3D texture, represented as RGB data.

```
vec3 world_to_uvw(vec3 position) {
    return (position.zxy + offsets) / detector_size;
}

vec3 Field(vec3 pos) {
    vec3 uv = world_to_uvw(pos);
    return textureLod(fieldTexture, uv, 0).xyz;
}
```

**Fig. 6.** Field access shader code for the Texture implementation.

if we do the same translation of the origin as in the previous case (see Fig. 4), and then divide the result by the dimensions of the detector (the detector_size global variable) — see world_to_uvw function in Fig. 6. We use the textureLod function to specifically ask for the textel of mipmap level 0, i.e. the original data (not mipmapped).

### 4.3. 3D sparse texture

In Fig. 5 it can be seen that a significant amount of space (around the detector) is occupied by zero vectors (represented by the black colour in the image). A lot of GPU memory can be saved if we find a way of not storing these regions. These savings could be in turn utilised by using a lower level of decimation, making the model more accurate for the same final storage cost. In OpenGL we can do this by using a special kind of texture called a Sparse Texture. It separates the reservation of GPU's address space (i.e. the texture size) from the requirement that all textures must be backed up by memory (commitment) and works in a similar fashion to virtual memory [20]. Writes to sparse texture's regions that are not committed yield no effect; with ARB_sparse_texture2 extension, reads from regions that are not committed yield zero data. The second property is particularly useful in our case, as we can just not commit detector regions where the field is zero — reads from those regions will yield the correct, zero vectors.

It is unfortunately not possible to control the memory commitment on a pixel level. The graphics card requires our program to operate on whole blocks of them — in case of graphics cards we tested the application on, $16 \times 16 \times 16$ pixel blocks. Due to this limitation an algorithm had to be created which would, based on the segment data from *AliROOT* (see Section 3), create a list of pixel blocks that should be committed in the sparse texture.

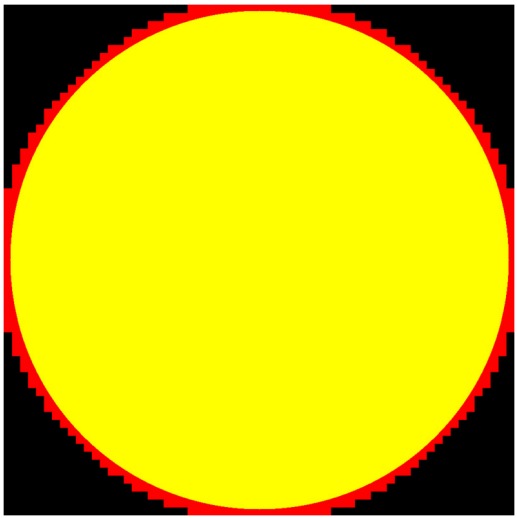

**Fig. 7.** Tight allocation of sparse texture blocks around the main detector chamber (frontal plane view). In red are the committed blocks of pixels. In yellow are positions that are both committed and have nonzero magnetic field.

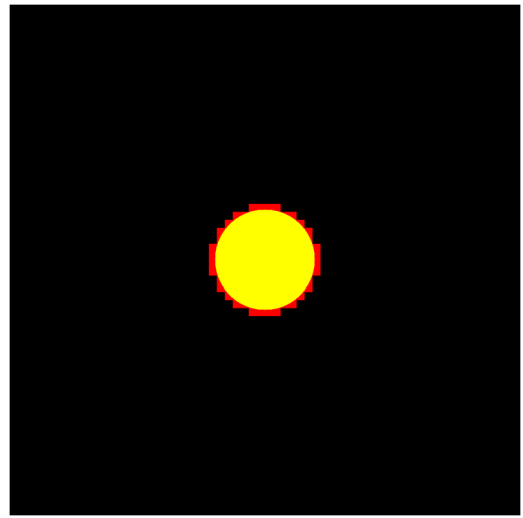

**Fig. 8.** Tight allocation of sparse texture blocks around the fragment of an LHC pipe (frontal plane view). In red are the committed blocks of pixels. In yellow are positions that are both committed and have nonzero magnetic field.

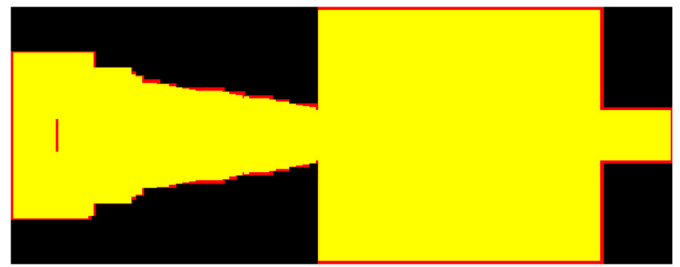

**Fig. 9.** Tight allocation of sparse texture blocks around the whole detector (transverse plane view). In red are the committed blocks of pixels. In yellow are positions that are both committed and have nonzero magnetic field.

An `std::unordered_set` was used to keep the list of blocks due to automatic removal of duplicates, which occur when a couple of *AliROOT* segments at least piecemeal share a single pixel block.

For the solenoid part of the detector we had concerns that floating point errors (when transforming back from cylindrical coordinate system of segments to Cartesian) might leave gaps in the block commitment. As the radius of the detector is known, we have decided instead to use a slightly modified Bresenham circle drawing algorithm, to "draw" a cylindrical shape tightly enveloping the actual detector (see Fig. 7).

In Fig. 5 (on the right side) it can be seen that the magnetic field model of the detector also includes a small portion of the **LHC** pipe, which can be handled by the same algorithm, just with a different radius parameter — see Fig. 8.

The dipole part of the detector, because it uses Cartesian coordinates, can be processed directly. For every dipole segment we fetch it's starting and ending position, then align the starting position to the pixel block grid (rounding down) and the ending position (rounding up), and then commit all blocks in space defined as such. That way the committed blocks envelop the segment, similarly to the solenoid case.

Overall, the committed blocks tightly envelop the whole detector (see Fig. 9), wasting as little of GPU memory as possible.

Using this technique we have managed to reduce the amount of utilised GPU memory by 49% (compared to the previous method — see Section 4.2). In the example, fourfold decimation case, this equals to around 250 MiB.

Because a sparse texture is handled like a normal texture inside the shader, this technique uses the same code as the previous method (Fig. 6) to access the field data.

### 4.4. GLSL

The model of the ALICE magnetic field can also be computed directly in shader code, in a similar fashion to the algorithm used in *AliROOT* (see Fig. 10). This significantly increases the computation complexity of the shader code, but provides almost perfect model accuracy (down to floating-point errors) and requires very little memory — segment definitions and Chebyshev polynomial parameters take 1.43 MiB and 1.02 MiB for the Dipole and the Solenoid part, respectively.

The original *AliROOT* code (see Section 3) uses cache — a global pointer to last used segment index — to skip the segment search if the current point requested by the user is inside the cached segment. Because the GLSL algorithm will be executed in parallel, such design, if implemented here, would cause a massive bottleneck.

For the GLSL implementation we propose a slightly different approach: the segment index can be kept in a buffer (array). Each shader invocation then operates on its own cell of the array, accessed using the primitive ID provided by the OpenGL in the shader code (either `gl_VertexID` in a vertex shader or `gl_PrimitiveIDIn` in a geometry shader). Memory for the array has to be allocated beforehand and filled with a known value. The actual segment IDs are non-negative, so we can choose -1 as an empty cache indicator. Due to memory allocation and initialization requirement the program needs to know in advance the maximum number of shader invocations (requested points), performed in a single batch.

We have also created a second version of the algorithm that does not use segment caching, so this extra memory (and its initialization) is not required. However, it might have potentially worse performance due to the segment search being always executed.

### 4.5. Constant

Lastly, to be able to easily compare the constant field method (see Section 1) with all the other developments, we have decided to implement it in our code as well. The constant approximation is only valid in the volume influenced by the Solenoid magnet. Boundaries of this volume can be easily checked once the input

```
vec3 SolDipField(vec3 pos) {
    if(pos.z>SOL_MIN_Z) {
        vec3 rphiz = CarttoCyl(pos);

        int segID = segment_cache.SolSegCache[gl_VertexID];

        if(segID >= 0 && IsInsideSol(segID, rphiz)) {
            vec3 brphiz = EvalSol(segID, rphiz);
            return CyltoCartCylB(rphiz, brphiz);
        }

        segID = findSolSegment(rphiz);
        if(segID >=0 && IsInsideSol(segID, rphiz)) {
            vec3 brphiz = EvalSol(segID, rphiz);
            segment_cache.SolSegCache[gl_VertexID] = segID;
            return CyltoCartCylB(rphiz, brphiz);
        }
    }

    int segID = segment_cache.DipSegCache[gl_VertexID];
    if(segID >= 0 && IsInsideDip(segID, pos)) {
        return EvalDip(segID, pos);
    }

    segID = findDipSegment(pos);
    if(segID >= 0 && IsInsideDip(segID, pos)) {
        segment_cache.DipSegCache[gl_VertexID] = segID;
        return EvalDip(segID, pos);
    }

    return vec3(0);
}

vec3 Field(vec3 pos) {
    if(pos.z > MIN_Z && pos.z < MAX_Z) {
        return SolDipField(pos);
    }
    return vec3(0);
}
```

**Fig. 10.** Field access shader code for the GLSL implementation.

```
const float SOL_MIN_Z = -550.f;
const float SOL_MAX_Z = 550.f;
const float SOL_MAX_R = 500.f;

vec3 CarttoCyl(vec3 pos) {
    return vec3(length(pos.xy), atan(pos.y, pos.x), pos.z);
}

vec3 Field(vec3 pos) {
    vec3 cyl = CarttoCyl(pos);
    if(cyl.z > SOL_MIN_Z && cyl.z < SOL_MAX_Z && cyl.x < SOL_MAX_R) {
        return vec3(0.0f, 0.0f, -5.0f);
    }

    return vec3(0.0f);
}
```

**Fig. 11.** Field access shader code for the Constant implementation.

coordinate is converted to cylindrical coordinates (see Fig. 11). Due to its simplicity, this implementation should be the fastest.

## 5. Experimental methodology

We have created two benchmark scenarios:

- *eval*, where we prepare a set of *N* points inside the detector volume and schedule the *AliROOT* or GPU implementation with fetching the field vector at each point,
- *fieldline*, where we prepare a set of *N*/100 points inside the detector volume and schedule the *AliROOT* or the GPU implementation to perform a simulated point drift – they should (1) fetch the field vector at a position, (2) apply its value to the position, then repeat from (1) 100 times. As the name of the benchmark implies, this operation creates a list of points along a particular field line of the magnetic field.

In both scenarios, the total amount of field value requests is equal to the (tunable) testing parameter *N*. We have run both benchmarks with the following values of *N*: 100, 200, 500, 1000, 2000, 5000, 10000, 20000, 50000, 100000, 200000. The number of steps per source point in the *fieldline* benchmark was arbitrarily chosen to be 100.

Three different sources of points were considered for the benchmark scenarios:

(a) randomly, uniformly generated from the whole detector volume,
(b) randomly, uniformly generated from the Solenoid magnet volume only,
(c) generated by track reconstruction software (points along the particle trajectories) from one of the Monte Carlo simulations available in the CERN database.

Because the constant field approach used in existing visualisation solutions (see Section 1) is valid only for the Solenoid magnet, the constrained generation (b) is used to obtain sensible results for the comparison. For the *eval* benchmark, random positions — (a) and (b) — can be used to prevent implementations from using their cache effectively, so the worst case scenario in terms of execution speed for each algorithm will be measured. The last source of positions (c) is used to evaluate the quality of the models in a similar way to a reconstruction task, i.e. sampled very densely in close proximity to the Interaction Point, and sparsely with increasing distance from the IP.

The *fieldline* benchmark scenario runs an operation that has a very similar access pattern to a physics track reconstruction (consecutive field evaluations in close proximity of each other), so the segment cache of tested algorithms should also be utilised in a similar way. As the starting point of the generated fieldline has no special considerations, we have used only the random Solenoid (b) source as the list of initial points, to be able to test all methods.

### 5.1. Retrieving data from the GPU

We have designed the benchmark in such a way that the resulting data from the computations should be available in computer memory. In *AliROOT*, which operates on the CPU, this is given. We need a way to fetch the results from the GPU memory for every other case.

For this task we have used the `Transform Feedback` feature of OpenGL, which allows vertices produced by a vertex or a geometry shader to also be written to an external buffer, that the application can then later retrieve. We are only interested in vertex processing, so we also disable rasterisation (via `glEnable(GL_RASTERIZER_DISCARD)` call), to stop the GPU from reaching further, redundant stages of the graphics pipeline.

For the *eval* benchmark, we use a vertex shader to fetch the field vectors; for the *fieldline* benchmark, we use a geometry shader to generate the position list from the initial point.

### 5.2. Execution time measurement

In both cases we measure the time that the computer spends doing actual computations, i.e. we exclude the initialization of the field model. To measure the elapsed time we have used the highest precision clock available in the standard library (`std::chrono::high_resolution_clock`).

In the case of GPU computations, we measure time spent between scheduling a new batch of computation (done by calling `glDrawArrays`) and finishing copying the results back to the CPU side (done by calling `glGetNamedBufferSubData`) — see Fig. 12.

```
glBeginTransformFeedback(GL_POINTS);

auto const auto start = std::chrono::high_resolution_clock::now();

glDrawArrays(...);
glEndTransformFeedback();

glGetNamedBufferSubData(...);

auto const end = std::chrono::high_resolution_clock::now();
```

**Fig. 12.** Time measurement schema for the GPU implementation.

```
auto const start = std::chrono::high_resolution_clock::now();

for (std::size_t i = 0; i < TOTAL_SAMPLES; ++i) {
    field->Field(points_in[i], points_out[i]);
}

auto const end = std::chrono::high_resolution_clock::now();
```

**Fig. 13.** Time measurement schema for the CPU implementation.

```
double RMSEx = 0.0, RMSEy = 0.0, RMSEz = 0.0;

for (std::size_t i = 0; i < TOTAL_SAMPLES; ++i) {
    glm::dvec3 v1 = mag->Field(points[i]);
    glm::dvec3 v2 = points_result[i];

    const auto diff = v1 - v2;

    RMSEx += diff.x*diff.x;
    RMSEy += diff.y*diff.y;
    RMSEz += diff.z*diff.z;
}
RMSEx = glm::sqrt(RMSEx/TOTAL_SAMPLES);
RMSEy = glm::sqrt(RMSEy/TOTAL_SAMPLES);
RMSEz = glm::sqrt(RMSEz/TOTAL_SAMPLES);
```

**Fig. 14.** Accuracy measurement schema which calculates the distance between points retrieved from the GPU and from the original model.

In the case of *AliROOT* computations, we measure how much time the CPU spends in the loop that processes requested points — see Fig. 13.

### 5.3. Accuracy measurement

To measure the field model accuracy against the ground truth (the *AliROOT* model) we have used the Root Mean Square Error for each individual axis — see Fig. 14.

In the *eval* benchmark, the RMSE is calculated on the field vectors. It is expressed in units of kGauss. In the *fieldline* benchmark, the RMSE is calculated on the positions of points in 3D space. It is expressed in centimetres.

## 6. Experimental results

Tests were performed on the following hardware:

- ThinkPad X1 Extreme — NVIDIA GeForce GTX 1050 Ti with Max-Q Design (4 GiB Video RAM), Intel Core i7-8850H 2.6 GHz.
- HPE ProLiant SL270s — NVIDIA Tesla K40m (12 GiB Video RAM), Xeon E5-2670 v2 2.5 GHz.

On the laptop computer, the benchmark was run on Ubuntu 20.04 (with NVIDIA Graphics Driver 460.39 installed). On the server machines the benchmark was run on the CentOS 6 Scientific Linux (with NVIDIA Graphics Driver 387.26 installed).

Each benchmark scenario was repeated 10 times to obtain a more representative sample. The results shown in the following sections are averages.

**Table 1**
RMSE values for experiments with maximum tested (200000) amount of points. Values are expressed in kGauss.

| Algorithm | RMSE (x) | RMSE (y) | RMSE (z) |
|---|---|---|---|
| Random points (detector volume) | | | |
| Shader Storage | 8.80e-01 | 1.83e-01 | 2.82e+00 |
| 3D Texture | 1.41e-01 | 4.24e-02 | 2.77e-01 |
| Sparse 3D Texture | 1.41e-01 | 4.24e-02 | 2.77e-01 |
| GLSL (nocache) | 3.67e-08 | 2.35e-08 | 5.38e-08 |
| GLSL (cache) | 3.67e-08 | 2.35e-08 | 5.38e-08 |
| Random points (Solenoid volume) | | | |
| Shader Storage | 1.16e-01 | 1.16e-01 | 4.91e+00 |
| 3D Texture | 3.92e-03 | 3.97e-03 | 6.19e-03 |
| Sparse 3D Texture | 3.92e-03 | 3.97e-03 | 6.19e-03 |
| GLSL (nocache) | 9.71e-05 | 7.48e-05 | 8.19e-05 |
| GLSL (cache) | 9.71e-05 | 7.49e-05 | 8.21e-05 |
| Constant | 1.16e-01 | 1.16e-01 | 2.38e-01 |
| Points from tracks | | | |
| Shader Storage | 4.50e-02 | 4.56e-02 | 4.90e+00 |
| 3D Texture | 1.85e-03 | 1.91e-03 | 3.56e-03 |
| Sparse 3D Texture | 1.85e-03 | 1.91e-03 | 3.56e-03 |
| GLSL (nocache) | 1.10e-08 | 1.06e-08 | 6.32e-08 |
| GLSL (cache) | 1.10e-08 | 1.06e-08 | 6.32e-08 |
| Constant | 4.50e-02 | 4.56e-02 | 1.87e-01 |

### 6.1. Benchmark eval

Table 1 presents the Root Mean Square Error measurements for the *eval* benchmark on the maximum amount (200000) of points tested for each scenario.

In both the random point and track point scenario, the requested point can land in an area where the field is zero, resulting in perfect accuracy. This can not happen if we consider the Solenoid magnet volume only. This is why, across the board, results for the Solenoid scenario are slightly worse than the other two.

In terms of errors, the worst implementation happens to be the Shader Storage Buffer. This approach in all cases is either worse, or on the same level as the Constant implementation. The texture approach (both standard 3D and Sparse) is in the middle, offering usually an order of magnitude better reflection of the original model than the Constant implementation. Unsurprisingly, the GLSL implementation of the *AliROOT* performs the best in every category. Its error rate is not zero most likely due to accumulation of floating point errors during the calculation. The original *AliROOT* algorithm operates on 64-bit variables, whereas GPU implementations operate on 32-bit variables. This design decision was made because consumer grade GPUs (on which the magnetic field visualisation will most likely be run) lack (or do not have enough instances of) fast double-precision floating point hardware. Even on Tesla GPU — considered as a high-end card — that we performed the experiment on, there are three times more standard cores than ones offering double-precision [17].

Figs. 15 and 16 present the total execution time of each scenario, with increasing number of test points *N*. To plot these graphs the distribution of points inside Solenoid volume was chosen due to being both uniform and supporting all implemented methods, especially Constant.

Here it can be seen that the Constant, Shader Storage Buffer and Texture implementations are very similar to each other in terms of speed and are the fastest. The Sparse Texture implementation is slower than those three, most likely due to not being fully hardware-accelerated. It can be seen that because of the randomly chosen points, the segment cache ends up being a net-negative for the GLSL method, performing slightly worse than the non-cache version. These versions managed to outperform the CPU

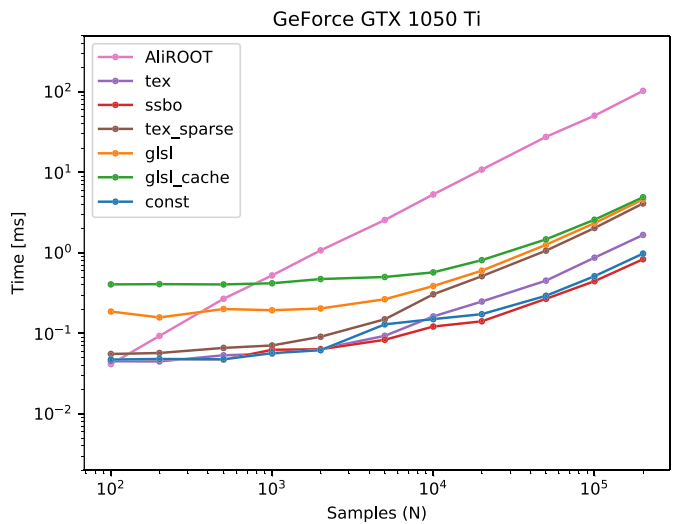

**Fig. 15.** Execution time (in milliseconds) of all implementations for the *eval* benchmark against the execution time of *AliROOT* algorithm, for each sample size.

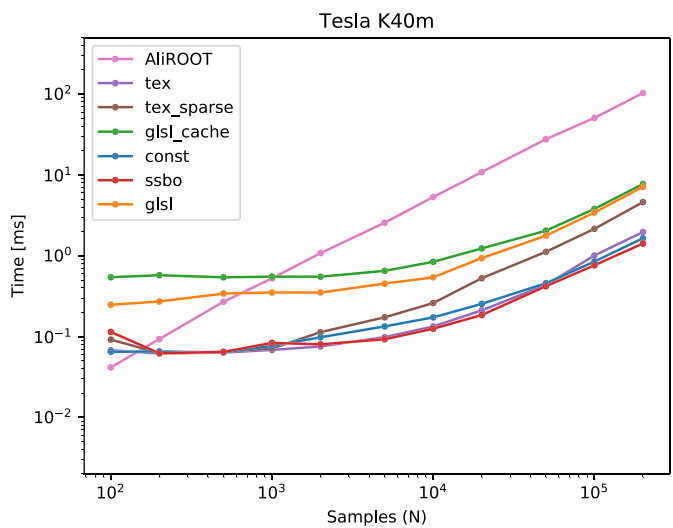

**Fig. 16.** Execution time (in milliseconds) of all implementations for the *eval* benchmark against the execution time of *AliROOT* algorithm, for each sample size.

implementation by the 1000 (cache) and 500 (no cache) requested points mark.

The Constant, Shader Storage Buffer and Texture implementations in the *eval* benchmark managed to complete the task faster than the *AliROOT* algorithm when asked for 200 points or more. Both GLSL implementations managed to do the same when asked for 1000 points or more.

### 6.2. Benchmark fieldline

Table 2 presents the Root Mean Square Error measurements for the *fieldline* benchmark on the maximum amount (200000) of points tested for each scenario. It should be noted that the due to the way points are generated this benchmark suffers from accumulating error, i.e. a slight discrepancy of the field value at a particular point will cause the remaining part of the fieldline path to be offset. This results in (relatively) larger errors than in the *eval* benchmark.

Regardless, the values follow similar trend to the *eval* benchmark. The GLSL implementations offer the best accuracy and the Constant implementation the worst. This time the Shader Storage Buffer performed similarly to the Texture approaches, all three of-fering an order of magnitude better accuracy than the Constant implementation.

**Table 2**
RMSE values for experiments with maximum tested (200000) amount of points (random distribution in the Solenoid volume). Values are expressed in centimetres.

| Algorithm | RMSE (x) | RMSE (y) | RMSE (z) |
|---|---|---|---|
| Shader Storage | 1.83e-02 | 1.85e-02 | 2.81e-02 |
| 3D Texture | 1.72e-02 | 1.75e-02 | 2.72e-02 |
| Sparse 3D Texture | 1.72e-02 | 1.75e-02 | 2.72e-02 |
| GLSL (nocache) | 9.27e-06 | 1.26e-05 | 1.03e-05 |
| GLSL (cache) | 9.27e-06 | 1.26e-05 | 1.03e-05 |
| Constant | 5.70e-01 | 5.54e-01 | 1.13e+00 |

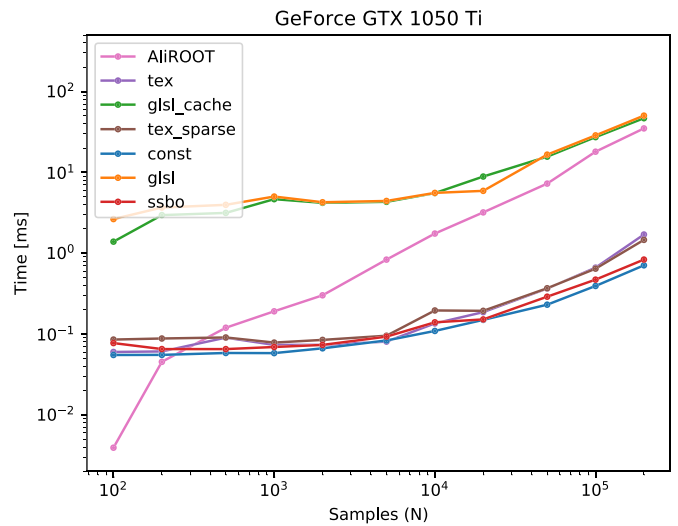

**Fig. 17.** Execution time (in milliseconds) of all implementations for the *fieldline* benchmark against the execution time of *AliROOT* algorithm, for each sample size.

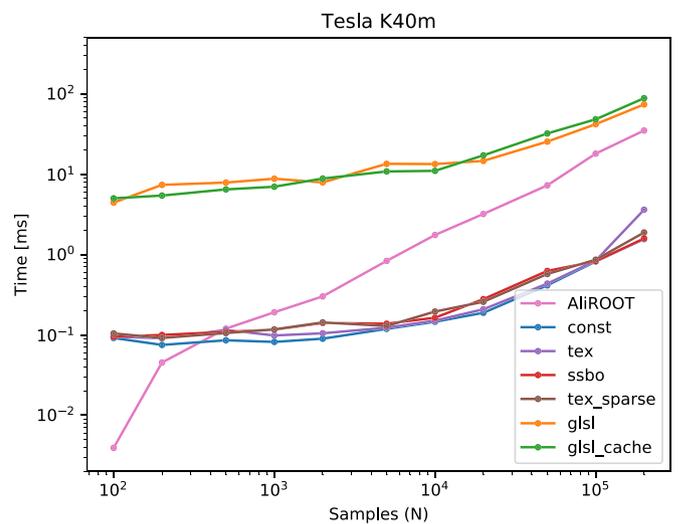

**Fig. 18.** Execution time (in milliseconds) of all implementations for the *fieldline* benchmark against the execution time of *AliROOT* algorithm, for each sample size.

Fig. 17 and 18 present the total execution time of each scenario, with increasing number of test points $N$. Here it can be seen that the repeated evaluation of the magnetic field in this benchmark causes significant loss of performance when GLSL implementations are used. These two approaches, unfortunately, were much slower than the CPU algorithm (around one order of magnitude slower in the worst case). This time the cache and non-cache version did not differ very much in execution speed.

On the other hand, the Shader Storage Buffer, Texture and Sparse Texture implementations perform very similarly to the *eval* benchmark. Because the *AliROOT* runs more efficiently this time, they managed to outperform it at slightly bigger sample size, at around 500 requested points.

## 7. Conclusion and future work

We have identified that the 3D visualisation tools (Event Display, MasterClass) used in ALICE experiment perform the track reconstruction using a very simple uniform field model. We have analysed how the detailed magnetic field algorithm (used in physics calculations) in the ALICE framework works and, based on this information, created six different implementations of the field model that can be run on the GPU for the purpose of visualisation.

We have created two benchmarks to test our implementations: a) *eval*, where we task the software with calculating field vectors at various points inside the detector volume; b) *fieldline*, where we task the software with generating field line vertices from various initial points inside the detector volume. In the test with the highest amount of points (200000) the *AliROOT* algorithm took 103 milliseconds to complete in the *eval* scenario. For comparison, the slowest implementation on the GPU (GLSL with cache) took 4.92 milliseconds on the 1050 Ti and 7.72 milliseconds on the K40m — around 13 – 20 times faster. These execution times are also equivalent to 129 – 203 frames per second.

In the same conditions in the *fieldline* scenario, the *AliROOT* algorithm took 35 milliseconds to complete. The GLSL (with and without cache) took 50 milliseconds on the 1050 Ti and 88 milliseconds on the K40m (20 and 11 frames per second), which is 42% – 150% slower. Other implementations performed much better — the Sparse Texture took 1.46 milliseconds (684 frames per second) on the 1050 Ti and 1.88 milliseconds (531.91 frames per second), so 18 – 23 times faster than the *AliROOT*.

Test results show that (with the exception of GLSL implementations of the original algorithm in *fieldline* benchmark), if enough work is scheduled for the GPU, the proposed techniques can be used to evaluate the field significantly faster than the *AliROOT* algorithm. Their performance is also sufficient for usage in a real-time rendering application, with the frame time margin large enough for other operations, such as processing of the field data. All implemented methods (except for the naive approach with the Shader Storage Buffer) offer 100 times or better accuracy than the uniform model (currently used in 3D applications).

The code used in the *fieldline* benchmark can be re-purposed to visualise the magnetic field of ALICE detector (see Fig. 19). Other ways of displaying the field data could be implemented following this approach, such as Line Integral Convolution [5], current tubes, current flows [21] and animations, to further increase the appeal of the visualisation.

Other possible direction of advancement is to re-implement the reconstruction of tracks used in Event Display and MasterClass for execution on the GPU as well, where the magnetic field models developed in this paper could be utilised for more accurate visualisation of tracks.

Lastly, the methods developed in this paper (with some adjustments) could also be used for visualisations of other detectors, as long as a similar Chebyshev polynomial–based model is used for their magnetic fields. One such example is the LHCb detector [15].

**Declaration of competing interest**

The authors declare that they have no known competing financial interests or personal relationships that could have appeared to influence the work reported in this paper.

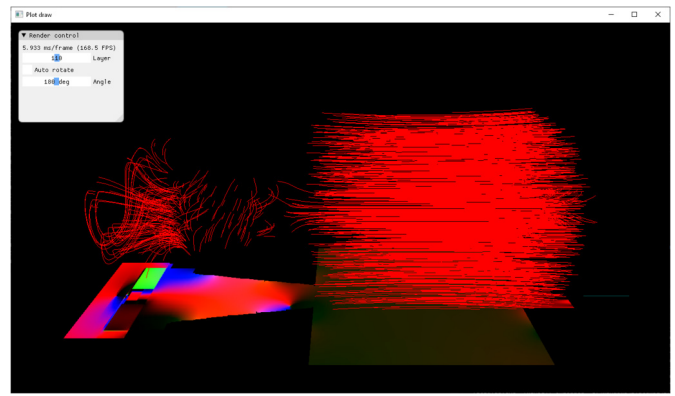

**Fig. 19.** An example of usage of the magnetic field data for visualisation.

**Acknowledgements**

We would like to thank the ALICE Collaboration for the support during our research as well as for the access to all software and internal design documents which were helpful in understanding the magnetic field algorithm.

This work was supported by the Polish Ministry of Education and Science under decision np. DIR/WK/2016/2018/17-1, Polish National Science Centre under decisions no. UMO-2016/22/M/ST2/00176, UMO-2016/21/D/ST6/01946, no. UMO-2017/27/B/ST2/01947, and by IDUB-POB-FWEiTE-1 project granted by Warsaw University of Technology under the program Excellence Initiative: Research University (ID-UB).

**References**

[1] K. Aamodt, et al., J. Instrum. 3 (2008) S08002, https://doi.org/10.1088/1748-0221/3/08/S08002.
[2] K. Adcox, et al., Nucl. Phys. A 757 (2005) 184–283, https://doi.org/10.1016/j.nuclphysa.2005.03.086.
[3] ALICE technical design report of the dimuon forward spectrometer, 8 1999.
[4] P. Buncic, M. Krzewicki, P. Vande Vyvre, Technical Design Report for the Upgrade of the Online-Offline Computing System, Technical report, CERN, Apr 2015, https://cds.cern.ch/record/2011297.
[5] B. Cabral, L.C. Leedom, in: Proceedings of the 20th Annual Conference on Computer Graphics and Interactive Techniques, SIGGRAPH '93, New York, NY, USA, Association for Computing Machinery, ISBN 0897916018, 1993, pp. 263–270.
[6] C. Cavicchioli, Development and Commissioning of the Pixel Trigger System for the ALICE Experiment at the CERN Large Hadron Collider, PhD thesis, Florence U, 2011.
[7] G. Dellacasa, et al., ALICE technical design report of the inner tracking system (ITS), 1999.
[8] G. Dellacasa, et al., ALICE time projection chamber: Technical Design Report, CERN-LHCC-2000-001, 2000, http://cds.cern.ch/record/451098.
[9] L. Evans, P. Bryant, J. Instrum. 3 (08) (aug 2008) S08001, https://doi.org/10.1088/1748-0221/3/08/s08001.
[10] P. Foka, M.A. Janik, Rev. Phys. 1 (2016) 154–171, https://doi.org/10.1016/j.revip.2016.11.002.
[11] P. Foka, M.A. Janik, Rev. Phys. 1 (2016) 172–194, https://doi.org/10.1016/j.revip.2016.11.001.
[12] Ł. Graczykowski, P. Nowakowski, P. Foka, EPJ Web Conf. 245 (2020) 08011, https://doi.org/10.1051/epjconf/202024508011.
[13] U.W. Heinz, J. Maurice, Evidence for a new state of matter: an Assessment of the results from the CERN lead beam program, 1 2000.
[14] J. Kapusta, B. Muller, J. Rafelski, Quark-Gluon Plasma: Theoretical Foundations, Elsevier, ISBN 978-0-444-51110-2, 2003.
[15] A. Keune, Nucl. Phys. B, Proc. Suppl. (ISSN 0920-5632) 197 (1) (2009) 163–166, https://doi.org/10.1016/j.nuclphysbps.2009.10.058, https://www.sciencedirect.com/science/article/pii/S0920563209007798, 11th Topical Seminar on Innovative Particle and Radiation Detectors (IPRD08).
[16] J. Niedziela, B. Haller, J. Phys. Conf. Ser. 898 (10 2017) 072008, https://doi.org/10.1088/1742-6596/898/7/072008.
[17] NVIDIA Kepler GK110 whitepaper, http://www.nvidia.com/content/PDF/kepler/NVIDIA-Kepler-GK110-Architecture-Whitepaper.pdf. (Accessed 15 August 2021), 2012.
[18] J. Rafelski, Eur. Phys. J. A 51 (9) (2015) 114, https://doi.org/10.1140/epja/i2015-15114-0.

[19] M. Richter, J. Phys. Conf. Ser. 664 (2015) 082046, https://doi.org/10.1088/1742-6596/664/8/082046.

[20] M. Segal, K. Akeley, The OpenGL® graphics system: a specification, https://www.khronos.org/registry/OpenGL/specs/gl/glspec46.core.pdf. (Accessed 15 August 2021), 2019.

[21] C. Stoll, S. Gumhold, H.-P. Seidel, in: VIS 05. IEEE Visualization, 2005, 2005, pp. 695–702.

[22] S. Yamasaki, Fast Magnetic Field Query Algorithm for the ALICE O2 Project, Master's thesis, Hiroshima University, Japan, 2018.